# Impact of Fountain Codes on GPRS channels

M Usman, J Dunlop
University of Strathclyde, Glasgow
{mohammed.usman, j.dunlop}@eee.strath.ac.uk

*Abstract* – The rateless and information additive properties of fountain codes make them attractive for use in broadcast/multicast applications, especially in radio environments where channel characteristics vary with time and bandwidth is expensive. Conventional schemes using a combination of ARQ (Automatic Repeat reQuest) and FEC (Forward Error Correction) suffer from serious drawbacks such as feedback implosion at the transmitter, the need to know the channel characteristics apriori so that the FEC scheme is designed to be effective and the fact that a reverse channel is needed to request retransmissions if the FEC fails. This paper considers the assessment of fountain codes over radio channels. The performance of fountain codes, in terms of the associated overheads, over radio channels of the type experienced in GPRS (General Packet Radio Service) is presented. The work is then extended to assessing the performance of Fountain codes in combination with the GPRS channel coding schemes in a radio environment.

## 1. Introduction

Fountain codes are rateless and information additive and are based on sparse graphs. They are rateless in the sense that an infinite sequence of encoded symbols can be potentially generated from a given sequence of message symbols. This means that the code rate need not be fixed before the transmission begins i.e. as many encoded symbols as needed to decode the complete set of message symbols may be generated. Fountain codes are information additive meaning that any subset of slightly greater than '$k$' encoded symbols are sufficient to decode the '$k$' message symbols that constitute the data being transmitted. This means that it does not matter which encoded symbols are received as long as sufficient of them are received, obviating the need for retransmission of lost or erroneous symbols and hence obviating the need for a reverse channel. These properties of fountain codes make them attractive for use over time-varying and bandwidth-limited radio channels, particularly for broadcast/multicast applications. Furthermore, they have fast encoding and decoding algorithms which make them attractive for time constrained applications such as streaming video.

Although fountain codes have been standardised for Multimedia Broadcast and Multicast Service, the literature on the performance of fountain codes over mobile radio channels is sparse. This paper aims to fill this void by assessing the performance of fountain codes over wireless channels of the type experienced in GPRS. Previous work [1] had considered the development of a testbed for the assessment of fountain codes over various types of channels and presented the performance of fountain codes over the Binary Symmetric Channel (BSC) and the Additive White Gaussian Noise (AWGN) channel. The testbed architecture described in [1] is used for the assessment of fountain codes over radio channels. The encoding and decoding algorithms for fountain codes are described in [2].

A reliable decoder for fountain codes is one which can recover the '$k$' message symbols using any $k' = k(1 + \varepsilon)$ encoded symbols. The factor $(1+\varepsilon)$ is called the decoding inefficiency. For good fountain codes $k'$ is close to $k$ i.e. $\varepsilon$ close to zero.

The first practical realisation of fountain codes are called as Luby Transform (LT) codes and are generated using the Robust Soliton (RSol) [2] distribution as their degree distribution. The degree of an encoded symbol is defined as the total number of message symbols used in its computation. The RSol distribution $\mu(d)$ is an enhancement of the Ideal Soliton distribution $\rho(d)$ and is defined as follows:
Let
$$R = c.\log_e(k/\delta)\sqrt{k}$$
for some constant $c > 0$. '$c$' is a free parameter with a value smaller than 1 giving good results. '$\delta$' is a bound on the probability that the decoding will fail to run to completion after a certain number $k' = k(1 + \varepsilon)$ encoded symbols have been received.

$$\rho(d) = \begin{cases} 1/k & \text{for } d = 1 \\ 1/d(d-1) & \text{for } d = 2,3......k \end{cases}$$

$$\tau(d) = \begin{cases} R/d.k & \text{for } d = 1,2.........(k/R)-1 \\ R.\log_e(R/\delta)/k & \text{for } d = k/R \\ 0 & \text{for } d = (k/R)+1..........k \end{cases}$$

$$\beta = \sum_{d=1}^{k} \rho(d) + \tau(d)$$

$\mu(d) = (\rho(d) + \tau(d))/\beta$  for $d = 1............k$

The RSol distribution ensures that there are sufficiently large number of encoded symbols having low degree so that the decoding process can get started and continue. At the same time it also ensures that there are occasional encoded symbols that have high degree so that all the message symbols are used in the encoding process with high probability.

## 2. Simulation Set-up

An important application of fountain codes being considered is the delivery of content over mobile radio channels. An assessment of the performance of fountain codes over such channels is therefore essential. The GPRS radio channel is modelled by incorporating data derived from a soft-decision Viterbi decoder [3]. The radio channel model used in this paper is that of a typical urban channel for a mobile travelling at 5km/hr. The simulations are performed for packet data transmissions in a GPRS like system for different SIR (Signal to Interference Ratio) values. The data to be transmitted is arranged as message symbols of fixed size of 32 bits each and transmitted over the radio channel in a Radio Link Control (RLC) block, which has a fixed size of 456 bits.

Fig 1 [4] depicts the steps involved in using the soft-bit magnitudes derived from the Viterbi decoder to model the radio channel. The data to be transmitted is mapped onto +/-1 depending on whether it is a binary one or a binary zero respectively. The mapped data is then multiplied with the appropriate soft bit magnitude. Instead of interleaving the data prior to multiplying with the soft bit magnitudes and then de-interleaving again following the multiplication, it is simpler to de-interleave the soft-bit magnitudes and then multiply the data with the de-interleaved soft-bit magnitudes [4]. The result of the multiplication is effectively the received data over the corresponding radio channel.

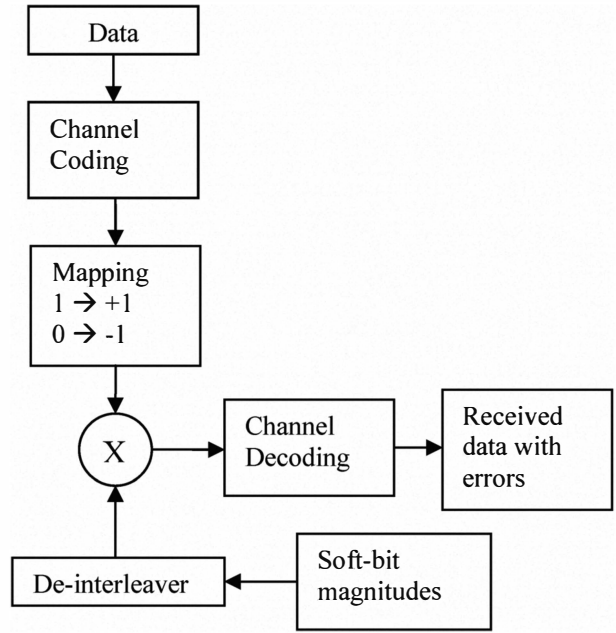

Fig 1 Schematic to model radio channels using soft bits

The received data is interpreted as a binary 1 if the product of the mapped data bit with the de-interleaved soft bit magnitude is positive and as a binary 0, if it is negative. In the first set of experiments where fountain codes are used without channel codes, the schematic of Fig 1 would not have the 'channel coding' and 'channel decoding' blocks.

## 3. Fountain codes over radio channels without channel coding

The data to be transmitted is fountain encoded and transmitted in a RLC block. The size of each fountain encoded symbol is the same as that of the message symbol i.e. 32 bits. Each RLC block therefore carries 14 fountain encoded symbols. For the sake of simplicity, it is assumed that errors in received RLC blocks are detected by some appropriate means (eg CRC) and the erroneous RLC blocks are simply discarded. The loss of a single RLC block therefore results in the loss of 14 fountain encoded symbols. The plot of fig 2 shows the fraction of message symbols recovered against the total number of fountain encoded symbols transmitted for various SIR values.

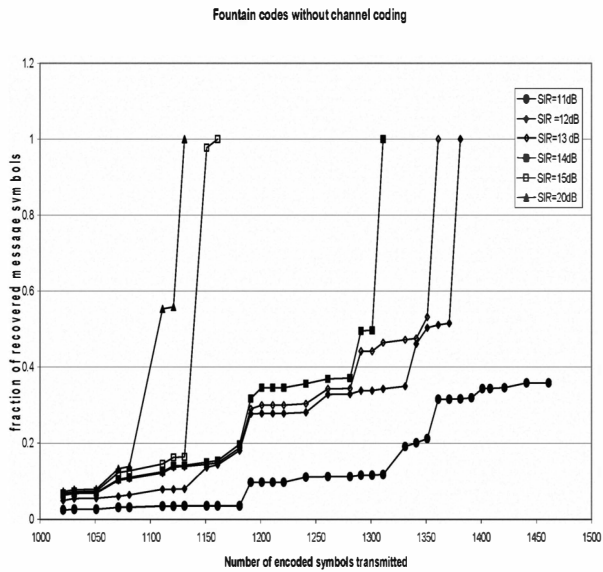

Fig 2 Fountain codes over radio channels without channel coding

The chain reaction property of fountain codes is clearly evident from these plots whereby the decoding initially proceeds gradually and then suddenly, a large number of message symbols are recovered. It is also clear that at lower SIR values, it takes much longer for the chain reaction to set in. This is expected as the losses are higher at lower SIR values. It is observed that for SIR values of 11 dB and below, the decoder fails to recover the complete set of '$k$' message symbols. Table 1 below summarises the percentage overheads for different SIR values. The overheads mentioned in this section refer to fountain overheads which actually represent the total overheads when channel coding is not used. Fountain overhead is the fraction of excess fountain encoded symbols over '$k$' which need to be transmitted in order to decode the '$k$' message symbols.

An important point to be noted here is that there were no retransmissions of symbols received in error. Erroneous symbols were simply discarded and not used in the fountain decoding process.

| SIR (dB) | % overhead |
|---|---|
| 20 | 10.7 |
| 15 | 13.7 |
| 14 | 28.4 |
| 13 | 33.3 |
| 12 | 35.26 |
| 11 | Decoding failure even after fountain overhead in excess of 45% |

Table 1 % overheads for fountain codes over radio channels at various SIR values

## 4. Fountain codes in combination with GPRS channel coding schemes

Instead of simply discarding the symbols received in error, error correction coding is applied to the fountain encoded data so that errors in the received RLC block are corrected. An RLC block is dropped only if it contains residual errors following channel decoding. In this paper, it is aimed to study the impact of fountain codes on existing data transmission systems such as GPRS. GPRS channel coding schemes are therefore chosen to be used in combination with fountain codes. Fig 3 shows the block diagram representing the complete processing performed on the data between the source and destination.

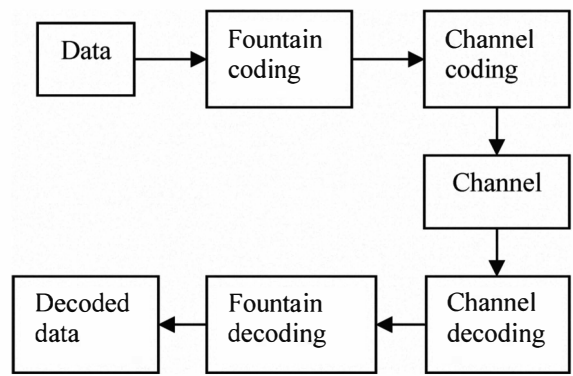

Fig 3 Block diagram of the processing performed on the data to be transmitted

GPRS defines four different coding schemes. The RLC block's data field length depends on the coding scheme used. The Uplink Status Flag

(USF) bits are fixed to zeros and ignored in the simulations presented in this paper. CS-1 uses a 40 bit Fire code and a ½ rate convolutional code for each RLC block. CS-1 is the most robust of the GPRS channel coding schemes and is generally used for RLC blocks carrying control information. CS-2, CS-3 and CS-4 use a 16 bit Block Check Sequence (BCS) for error detection. A ½ rate convolutional code followed by puncturing is used for error correction in CS-2 and CS-3 resulting in a blocks of 456 coded bits. CS-4 has no error correction. Table 2 summarises the four GPRS channel coding schemes [5].

| Code | Code Rate | Pre-Code USF | Data Bits | BCS | Tail | Coded bits | Punctured bits |
|---|---|---|---|---|---|---|---|
| CS-1 | 1/2 | 3 | 181 | 40 | 4 | 456 | 0 |
| CS-2 | ≈2/3 | 6 | 268 | 16 | 4 | 588 | 132 |
| CS-3 | ≈3/4 | 6 | 312 | 16 | 4 | 676 | 220 |
| CS-4 | 1 | 12 | 428 | 16 | - | 456 | 0 |

Table 2 GPRS coding schemes

The performance of fountain codes in combination with GPRS coding schemes CS-1 CS-2, CS-3 and CS-4 are presented in this paper. Since CS-4 has no error correction, the performance of fountain codes in combination with CS-4 would be the same as that presented in section 3. It must also be noted that link adaptation is not used in any of the results presented. The performance of fountain codes in combination with the individual coding schemes is assessed for different SIR values.

## 5. Assessment of Fountain Codes in combination with Coding Schemes CS-3 and CS-2

The data to be transmitted is organized as symbols of 32 bits each. Fountain codes do not lay any restriction on the size of symbols. A symbol can be a single bit or any number of bits. A value of 32 bits/symbol is chosen for ease of representation of the symbols within the simulation environment. In this paper, the symbols representing the data to be transmitted are referred to as the message symbols and their total number is referred to as '$k$'. Fountain encoded symbols are generated using these message symbols with the Robust Soliton distribution as the degree distribution.

Channel coding scheme CS-3 is then applied and the resulting RLC blocks are interleaved as specified in [6] and transmitted over the simulated radio channel at various SIR values. The interleaving and de-interleaving is performed as described in section 2. The convolutional code is decoded with a soft-decision Viterbi decoder followed by the decoding of the BCS. If the BCS for an RLC block fails, the complete block is discarded and the fountain encoded symbols it carried are considered to be lost. No requests for retransmissions are made. The decoder simply takes in new blocks and extracts the fountain encoded symbols within the block only if the block is error free following channel decoding. Fountain decoding is applied once '$k$' fountain encoded symbols have been received. If fountain decoding fails to recover all the '$k$' message symbols, more RLC blocks are transmitted and hence more fountain encoded symbols and fountain decoding applied again. Fountain decoding is terminated if it fails to recover all the '$k$' message symbols even after transmission of large number of fountain encoded symbols. The simulation is then repeated by applying channel coding scheme CS-2. The plots of figure 4 and figure 5 show the performance of fountain codes in combination with CS-3 and CS-2 respectively.

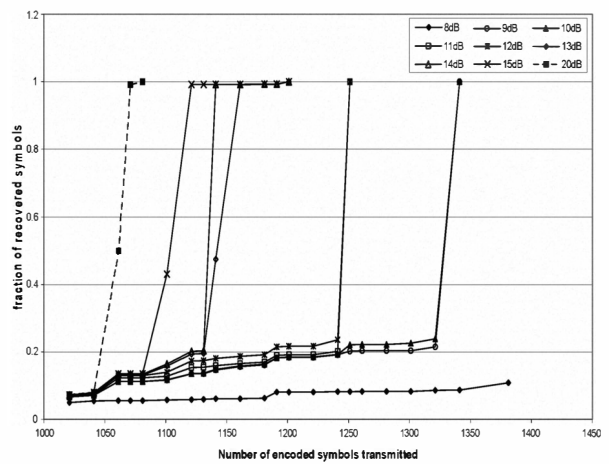

Fig 4 Fountain codes with channel coding scheme CS-3

A comparison of the plot in figure 4 with that of figure 2 reveals that the overheads in terms of the total number of fountain encoded symbols needed to be transmitted (called fountain overheads) in order to decode the complete set of message symbols is lower when CS-3 is used than when CS-4 is used (no channel coding). Furthermore, with CS-3 as the channel coding scheme, fountain decoding fails to decode all the '$k$' message symbols when the SIR drops to 8 dB and lower. The corresponding threshold when CS-4 is used is 11 dB.

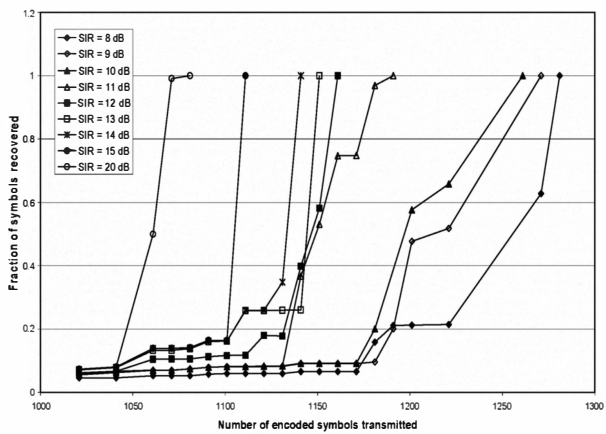

Fig 5 Fountain codes with channel coding scheme CS-2

A comparison of the plots in figures 2, 4 and 5 reveals that fountain overheads are lower when CS-2 is used in combination with fountain codes than when CS-3 or CS-4 are used. It is also observed that with CS-2, decoding is successful for SIR values as low as 8 dB.

## 6. Assessment of Fountain Codes in combination with Coding Scheme CS-1

Fountain encoded symbols are generated as before by drawing the degrees from the Robust Soliton distribution. Channel coding scheme CS-1 is then applied to the fountain encoded data followed by interleaving as specified in [6] and then transmitted over the simulated radio channel for various SIR values.

The received RLC blocks are channel decoded by first decoding the convolutional code with a soft decision Viterbi decoder as before followed by the decoding of the Fire code. If the BCS for an RLC block fails, it is discarded. Again, no requests for retransmissions are made and there are no retransmissions of discarded RLC blocks. Fountain encoded symbols are extracted from the error free RLC blocks following channel decoding and then fountain decoding is performed. The plot of figure 6 shows the performance of fountain codes when CS-1 is used for channel coding.

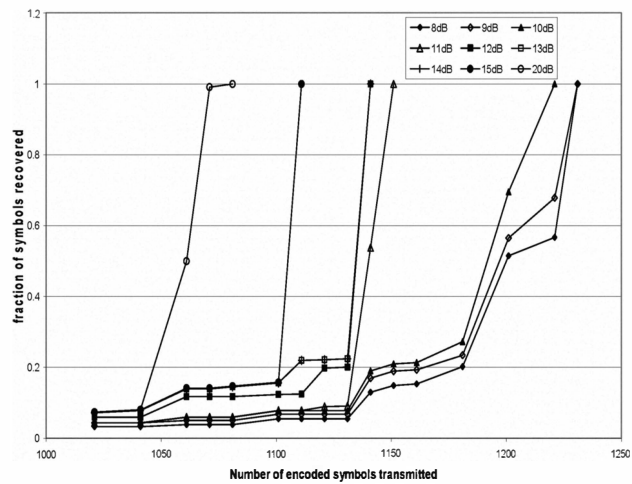

Fig 6 Fountain codes with channel coding scheme CS-1

Comparing the plots figures 2, 4, 5 and 6, it becomes clear that fountain overheads are smallest when CS-1 is used and largest when CS-4 is used in conjunction with fountain codes. It is also observed that the gains achieved by each of the coding schemes, in terms of reduced fountain overheads reach a stage of diminishing returns as the channel SIR is increased.

With CS-1, the fountain decoder successfully decodes all the '$k$' message symbols for SIR values as low as 8 dB with a fountain overhead of 20.57%. The significance of this improvement can be easily appreciated when considering that fountain overheads in excess of 35% are encountered for a SIR value of 12 dB when channel coding scheme CS-4 is used. Furthermore, fountain decoding fails to recover all the '$k$' message symbols even with large fountain overheads, for channels with SIR values lower than 12 dB when CS-4 is used. It must be noted, however, that fountain overheads do not include overheads associated with channel coding. The total overheads (fountain overheads + channel coding overheads) are discussed in section 7.

Table 3 summarises the percentage fountain overheads for the four cases i.e. fountain codes in combination with coding schemes CS-4 (i.e. no channel coding), CS-3, CS-2 and CS-1.

| SIR(dB) | % Fountain overheads | | | |
|---|---|---|---|---|
| | CS-4 | CS-3 | CS-2 | CS-1 |
| 8 | 43.09 * | 43.09 * | 25.46 | 20.57 |
| 9 | 43.09 * | 31.34 | 24.49 | 20.57 |
| 10 | 43.09 * | 31.34 | 24.49 | 19.58 |
| 11 | 43.09 * | 22.53 | 16.65 | 12.73 |
| 12 | 35.25 | 22.53 | 13.71 | 11.75 |
| 13 | 33.30 | 17.63 | 12.73 | 11.75 |
| 14 | 28.40 | 17.63 | 11.75 | 11.75 |
| 15 | 13.71 | 11.75 | 8.81 | 8.81 |
| 20 | 10.77 | 5.88 | 5.88 | 5.88 |

Table 3 Comparison of fountain overheads

## 7. Total Overheads associated when using Fountain Codes in combination with GPRS Channel Coding Schemes

The overheads described in sections 5 and 6 refer only to the fountain overheads i.e. the fraction of excess fountain encoded symbols that need to be transmitted in order to decode a set of '$k$' message symbols. The channel coding schemes introduce additional overheads and these need to be accounted for. The plot of figure 7 shows the total overheads associated for the four simulation scenarios i.e. fountain codes with channel coding schemes CS-4, CS-3, CS-2 and CS-1, at various SIR values.

In this section, the total overheads (fountain overheads + channel coding overheads) involved are presented. In the case of fountain codes with CS-4, the total overheads are the same as the fountain overheads, as stated in section 3. An observation of the plot in figure 7 makes it clear that the total overheads are substantially larger when channel coding is applied, although the use of channel coding results in a reduction in fountain overheads, as indicated in table 3.

---

* in Table 3 indicates that fountain decoding failed to recover all the '$k$' message symbols at those SIR values

It is clear from figures 2 and 4 that channel codes CS-4 and CS-3 are not suitable for use in conjunction with fountain codes when the channel SIR drops below 12 dB and 9 dB respectively. In these circumstances, fountain decoding fails even with considerably large fountain overheads and would not fulfill the requirements for multicast. The plots of figures 5 and 6 indicate the advantage of using CS-2 and CS-1 when the SIR is as low as 8 dB, both of which allow successful decoding of the complete set of '$k$' message symbols with lower fountain overheads.

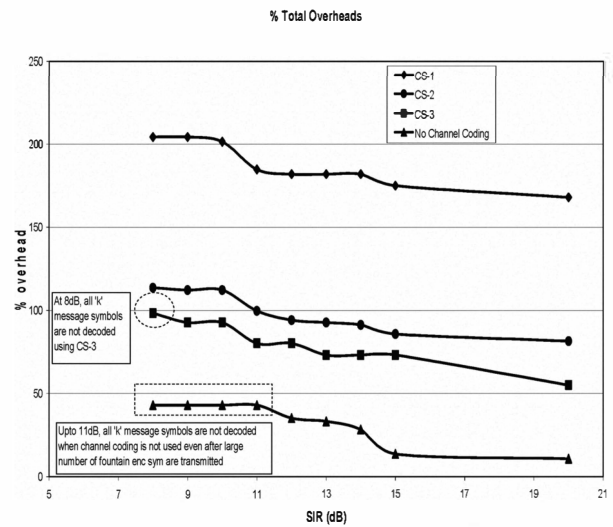

Fig 7 Total overheads

However, it is clear from the plot of figure 7 that both CS-1 and CS-2 have considerably larger additional overheads associated with them. The points surrounded by dotted lines in figure 7 indicate decoding failures. Furthermore, the use of CS-1 in conjunction with fountain codes would be wasteful except under very poor channel conditions as complete recovery of the message symbols is achieved with the use of CS-2, which has much lower total overheads as compared to CS-1. Therefore, under these circumstances, some form of link adaptation would be appropriate, using channel coding scheme CS-3 when the SIR drops below 12 dB and using coding scheme CS-2 when the SIR drops below 9 dB. When the channel SIR is above 15 dB fountain codes are able to perform with minimum overheads.

Under poor channel conditions, most of the RLC blocks are received in error resulting in them being discarded. However, with channel coding, a RLC block is discarded only in the presence of residual errors following error correction. Thus the total number of RLC blocks and hence the total number fountain encoded symbols discarded due to errors in the received blocks is lower when channel coding is used than otherwise.

## 8. Conclusion

The performance of fountain codes over GPRS channels has been evaluated for various SIR values. A comparison is made between fountain codes in combination with the four different GPRS channel coding schemes. Although the use of channel coding results in a reduction of fountain overheads, the channel codes themselves introduce considerably large additional overheads. For channels with SIR value above 15 dB, fountain codes are able to perform with minimum overheads. However, under poor channel conditions, the use of channel coding makes it possible to completely recover the original message symbols. This is not the case when channel coding is not used where a majority of the received blocks of data are found to be in error and therefore discarded.

Thus, it can be concluded that the use of fountain codes in combination with GPRS channel coding schemes would be appropriate with some form of link adaptation. It is however, necessary to study the performance of fountain codes in a typical radio environment (dynamic channel conditions), to determine the point at which the transmitter can stop transmitting and still ensure that the receiver will have received sufficient encoded symbols required to decode the '$k$' message symbols with high probability. This will help fully utilise the important advantage offered by fountain codes i.e. obviating the need for a reverse channel. The application of link adaptation under varying channel characteristics and a comparison with an acknowledgement based scheme is also necessary in order to understand the impact of fountain codes in a more practical scenario. These will be the subject of further work.